# Reconstruction of the evolutionary history of gene gains and losses since the last universal common ancestor

Haiming Tang[1,*]; Paul Thomas[1,2]; Haoran Xia[3];

1 Department of Preventive Medicine, University of Southern California, Los Angeles, California, USA

2 Department of Biological Sciences, University of Southern California, Los Angeles, California, USA

3 Department of Earth Sciences, University of Southern California, Los Angeles, California, USA

*Corresponding author

**Abstract**:

*Gene gains and losses have shaped the gene repertoire of species since the universal last common ancestor to species today. Genes in extant species were gained at different historical times via de novo creation of new genes, duplication of existing genes or transfer from genes of another species (HGT), and get lost gradually. With the increasing number of sequenced genomes, some comparative analyses have been done to quantify the evolutionary history of gene gains and losses in restricted lineages like vertebrates, insects, fungi, plants and so on. Here, we have constructed and analyzed over 10,000 gene family trees to reconstruct the gene content of ancestral genomes at an unprecedented scale, covering hundreds of genomes across all domains of life. This is the most comprehensive genome-wide analysis of all events in gene evolutionary histories. We find that our results are largely consistent with earlier, less complete comparative studies on specific lineages such as the vertebrates, but find significant differences especially in recent evolutionary histories. We find that the rate of gene gain varies widely among branches of the species tree, and find that some periods of rapid gene duplication are associated with great extinctions in geological history.*

# Introduction

The evolutionary history of gene repertoires at different historical periods is an intriguing topic. The gene repertoires of species must have changed dramatically from the earliest form of life: the last universal common ancestor (LUCA) about 4200 million years ago to complex organisms that are present today (Darwin Charles 1859; Penny and Poole 1999; Glansdorff et al. 2009; Weiss et al. 2016).

Through comparative analysis of genomes of extant species, we could find homologs that share similar sequence, structure and function, which are believed to have evolved from a shared common ancestor. This also yields reconstruction of different types of events in gene evolutionary history: new genes could arise through gene duplication by copying pre-existing genes in the same genome, de novo creation (Kaessmann 2010; McLysaght and Guerzoni 2015) or through horizontal transfer which is a transfer of genes from genome of a different species (Gogarten and Townsend 2005; Keeling and Palmer 2008; Soucy et al. 2015); genes could get lost through psedudogenization, leads to immediate loss of gene function (Wang et al. 2006; Albalat and Canestro 2016).

The birth of new genes has been recognized as a major force for emergence of new functions and adaptive evolution of species (Zhang et al. 2001; Zhang 2017). Most of the new genes have come from duplication of previous homologous genes instead of from de novo creations. Tandem duplication refers to duplication of exons within the same gene which gives rise to the subsequent genes. It usually displays as occurrence of two or more similar or identical sequences. A complete exon analysis of all genes in *Homo sapiens*, *Drosophila melanogaster*, and *Caenorhabditis elegans* has shown a total of 12,291 instances of tandem duplication in exons of these 3 species (Letunic et al. 2002). Tine Bloome et al. has found extra-large number of duplications at the time periods just before divergence of the vertebrates, which are much larger than estimated duplications at later time periods like tetrapod, and mammals (Blomme et al. 2006). Zhou et al. studied gene duplications in early eukaryotes through gene family reconstruction of 14 species in the animals, fungi and plants kingdoms (Zhou et al. 2010). Some other comparative studies have focused on specific lineages such as the primates (Hahn et al. 2007), insects (Wyder et al. 2007), colletotrichum genus (Gan et al. 2016), and reconstructed gene events at these specific phyla.

The ''less-is-more'' hypothesis proposed gene loss as an engine of evolutionary change. Studies have shown the impact of gene losses in shaping the human lineage through comparison of human pseudogenes with a putatively functional chimpanzee ortholog. Gene loss has been recognized as a pervasive source of genetic variation that can cause adaptive phenotypic diversity (Albalat and Canestro 2016).

A diversity of questions could be addressed by analysis of gene evolutionary events and the inferential power for understanding gene functions. Evolutionary history of gene gains and losses are informative about genes function (Capra et al. 2013). The concept of "gene age" is frequently equated with the timing of evolutionary events: the ceancestorspecies associated with duplication, horizontal gene transfer and de novo creation events. A gene which traces to a duplication event at Opistokonts is "older" than a gene which traces to a duplication event at a ceancestorthat appears in later ceancestors such as Vertebrates. Gene's ages are found to be closely related with when genes are expressed during embryonic development. Species in many phyla progress through a "phylotypic" stage, in which species with highly divergent adult morphologies display dramatic phenotypic similarities. Gene expression study in the development of zebrafish, flies, and nematodes demonstrated that genes expressed in the phylotypic stage are significantly "older" than those expressed in earlier and later developmental stages that show species-specific characteristics (Domazet-Loso and Tautz 2010). Protein Interaction Network in several yeast species are found to be dependent on age of yeast genes: "younger" genes are found to have fewer physical interactions compared with "older" genes (Kim and Marcotte 2008). Gene's evolutionary history is also tightly related with its functionality in diseases. Most cancer related genes are found to be dated to two evolutionary periods: the age for cellular organisms and the age of metazoan when multicellular animals emerge (Domazet-Loso and Tautz 2008). It was also found that genes associated with Mendelian diseases are "older" than others, including genes that are associated with complex diseases (Cai *et al*. 2009). Our previous study has also shown the validity of preserved ages of amino acids for predicting deleterious variants (Tang and Thomas 2016).

Advances in sequencing technologies are rapidly making complete nucleotide sequence of organisms routinely available. The abundance of available genomes has enabled a systematic comparative study of species from all major phyla in the tree of life. To our knowledge, this study reconstructs for the first time a most comprehensive evolutionary history of gene gain and loss events since the universal common ancestor.

As described in detail in Methods, we have taken a comprehensive phylogenetic approach to reconstruction of protein-coding genomes, for common ancestors that humans share with 213 other organisms with fully-sequenced genomes across the tree of life. In previously published work (Mi, et al. 2017), we constructed nearly 15,000 phylogenetic trees, each of which is fully "reconciled" to the consensus species tree (Boeckmann, et al. 2015; Fang, et al. 2013). Here, we use these reconciled trees to infer the genes that were present in each ceancestor of the 213 species, and reconstruct the history of each gene from its first appearance, to subsequent inheritance and loss in descendant genomes. Our reconstruction covers ~95% of the known protein coding genes in the human genome (with the remaining 5% having evolved later than the last ceancestorof the great apes). We inferred gene evolution events at each ceancestor in the evolutionary history, either gained as a new gene (relative to its previous ancestor) by either gene duplication, de novo origination, or HGT or failed to inherit from ancestor by gene loss. The reconstructions reviewed significant differences of evolutionary histories for different lineages.

## Materials and Methods

### Gene families and reference taxonomic tree.

Gene trees were taken from the PANTHER database (Mi, et al. 2017) (www.pantherdb.org), SuperPANTHER version 1 (release date July 2015), though we also performed analyses with PANTHER version 12 (release date July 2017), with similar results. PANTHER trees are constructed using latest gene tree construction algorithm(Szollosi, et al. 2015). In summary, trees of gene families are constructed using 1) an assumed reference species tree, 2) knowledge of all recognizable members of a given family in each genome, 3) protein sequence data for each member, and 4) identification of potentially problematic gene predictions. A gene tree is constructed stepwise by creating "orthologous subtrees" that are connected via inferred duplication and transfer events. Each subtree contains orthologous sequences related by speciation events with at most one gene from each organism, and the tree topology is determined by the known species tree; a duplication event is inferred only when there is genomic proof that a duplication occurred, *viz*. when, during the iterative process, a given subtree contains more than one gene from the same species (within-species paralogs); a horizontal transfer event is considered if the number of gene deletions that would be implied by a history of vertical inheritance is too large. The tree is reconciled with the species tree, meaning that each internal node is labeled with the event type (speciation, duplication, transfer), and speciation nodes are labeled with the cenancestor from the species tree. For this study, we have used the largest version of PANTHER which contains 11,962 PANTHER™ families built from 214 genomes consisting of a total of 1,783,278 genes.

The reference species tree SuperPANTHER utilizes the consensus tree determined from an extensive review of the available literature(Boeckmann, et al. 2015). It is available in newick format in Supplemental file 1. The extant species are represented by the 5 alphanumeric characters as used in Uniprot Taxonomy database (https://www.uniprot.org/help/taxonomy).

The dating of ceancestors are from the single "Estimated Time" of divergence from the TimeTree resource (Kumar, et al. 2017).

### Phylogenetic analysis of gene evolutionary events.

Each gene tree is compared with the reference species tree to infer four types of gene evolutionary events that occurred along each branch of the tree, assuming the most parsimonious scenario. The four types of gene evolutionary events are gain by duplication ($D$), gain by horizontal transfer ($H$), gain by de novo origination ($R$) and gene loss ($L$). An example gene tree and species tree are illustrated in Figure 1. The gene gain events are readily available from internal node types: for duplication nodes, we first find its direct descendants (PANTHER trees may be unresolved such that a node has more than two descendants), and when total number of descendants is $n$, we can infer that a minimum of and $n$-1 duplication events must have occurred. For horizontal transfer nodes, we first find its direct descendants, and identify which of the descendants was vertically inherited by finding the first speciation node prior to the event, and determining which of the descendants of the transfer event is also a descendant in the reference species tree. The other descendant is considered to be transferred. If there is no ancestral speciation node in the tree, the event is considered to be unresolved, as we cannot infer in which branch of the species tree the gene was created, and to which branch it was transferred. For de novo gene origination, we assume one such event to have occurred per gene tree (origination of the root of the tree), and our task is to determine during what branch of the species tree the event occurred. If the root of the tree is a speciation event, we count the gene as originating along the branch prior to that event. If the root is a duplication event, we find the most ancient speciation event among its descendants (not necessarily immediate descendants). As mentioned above, if the root is a horizontal gene transfer event, as we are unable to infer which descendant is transferred and which is vertically inherited, we count it as unresolved.

Gene losses are estimated through comparison of gene tree structure with the reference species tree. The set of descendant lineages is compared with the set of descendant lineages of a corresponding common ancestral species. If the two sets are identical, then the most parsimonious scenario is that no gene loss occurred and we assume no gene loss. If a gene is missing from descendant lineages in the gene tree that are present in the species tree, we infer that gene loss occurred. We find the most parsimonious scenario (fewest losses) that explain the missing genes. If a lineage in the set of the gene tree structure is a descendant of a lineage in the set of the species taxonomy, then we will look recursively into the descendants of the lineage in the set of the species taxonomy until we find the lineage in the set of the gene tree structure. In this process, we count a loss in all descendant lineages of the species tree that do not overlap with the lineage in the set of the gene tree structure (Figure 1).

The software package for inferring gene evolutionary gene histories of PANTHER™ gene families is available at https://github.com/haimingt/genome_history_inference. For general applicability of comparing gene trees with reference species tree, we have also written the algorithm in format of pseudo codes. It is available in Supplemental file 2.

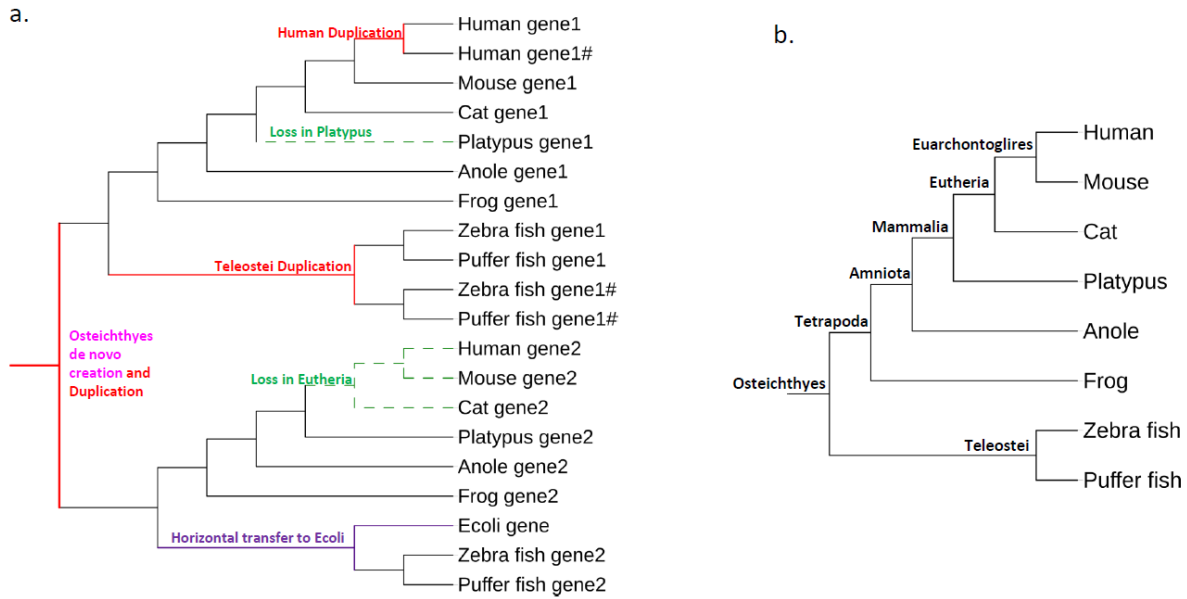

*Figure 1. Comparing a gene tree with the species tree to infer evolutionary events: D(Duplication)/L(Loss)/R(De novo)/H(Horizontal transfer).*

*This figure shows an example gene tree and species tree to illustrate the protocol to infer gene gains and losses for ancestral branches. On the right (b) is a species tree showing the relationships between several vertebrate species. The internal nodes of the species tree are labeled with the ceancestors. The cenancestor of all these vertebrate species is Osteichthyes (bony fish). On the left (a) is a hypothetical gene tree. The gene tree structure is reconciled with the species tree, inferring speciation events, duplication events and horizontal gene transfer events in the internal nodes. These events are analyzed to get branch specific statistics: gain by duplication (D), gene loss (L), gain through de novo origination (R), gain through horizontal transfer (H) for each ancestral species (branches in reference species tree). In this example, we will count one de novo origination prior to Osteichthyes; one duplication prior to Osteichthyes, Teleostei and Human; one loss prior to Eutheria and Platypus, and one horizontal transfer to the branch prior to E. coli.*

## Phylogenetic tree visualization

The phylogenetics trees are plotted using the iTol services (Letunic and Bork 2016). Pie charts are plotted on each internal branch and besides tree leaves. The slices of the pie charts represent the number of gene gains and losses by different sources. The size of the pie charts is proportional to the total number of gene copies gained and lost.

# Results and Discussion

In this study, we have constructed a complete evolutionary history of gene gains and losses, using SuperPANTHER verison 1.0, the PANTHER version which contains the largest number of genomes, thus yielding the most comprehensive reconstructions across all major phyla of life. The results are visualized on the tree of life in Figure 2.0. The raw numbers of gene gains and losses for each ceancestor are included in Supplemental file 3. We have also compared with earlier PANTHER version, and get similar results (PANTHER version 12.0 release date July 2017; the results are in supplemental file 4 and 5).

SuperPANTEHR 1.0 builds 11000 gene families from genomes of 214 species. Thorough our reconstructions, we build the gene evolution events for all major evolutionary stages, represented by in 153 ceancestors. Each ceancestor represents a common ancestor of a group of extant species that had existed in evolutionary history. For example, LECA, the last eukaryotic common ancestor is the common ancestor of all eukaryotes today which is hypothesized as an anaerobic and unicellular organism lived 2.1 billion years ago. We have formalized four types of gene evolutionary events: gain by duplication (D), gain by horizontal transfer (H), gain by de novo origination (R) and gene loss (L). Pie charts are plotted on the species tree for each internal node which represent ceancestors and leaves for extant species. The size of the pie circle is proportional to the total number of gene evolutionary events: gain by duplication (red), loss (green), gain by horizontal transfer (blue) and gain by de novo creation (magenta). We have collapsed several clades including Eubacteria, Archaea, Excavates and Alveolata-Stramneopiles due to the small number of species within these clades. Specifically for Prokaryotes, attempts to estimate the true number of prokaryotes diversity have ranged from $10^7$ to $10^9$ total species (Curtis, et al. 2002), while there are only 70 of species in the species tree, thus would unbalance the conclusions in evolutionary history of gene gains and losses in these clades.

The gene evolution reconstructions across all major phyla of life reveal lots of novel patterns. To show the validity of the reconstructions from the PANTHER database. We first compare with previous studies which were restricted on smaller phyla, like vertebrates. We then examined the deep evolutionary path of human evolution from the universal common ancestor. We also relate the massive dupliation periods in evolutionary history to great extinction events in geological history. We also examined the horizontal gene transfers among all major phyla for the first time. Finally, we discuss the novel unreported patterns of gene gains and losses and some potential artifacts.

## Gene evolutionary events in vertebrates

Tine Blomme *et al*. reconstructed gene duplications and losses by comparative analysis of 7 different vertebrate genomes: human, mouse, rat, chicken, frog, zebrafish and pufferfish (Blomme et al. 2006). They identified 2972 duplications in the branch before divergence of the ceancestor of these 7 vertebrates, and 545 duplications in the branch before divergence of zebrafish and pufferfish. The large number of duplications provide support for the hypothesis of 2 whole genome duplications (WGDs) before divergence of vertebrates (Kasahara 2007) and an additional fish specific genome duplication (FSGD) after fish's divergence from land vertebrates (Glasauer and Neuhauss 2014). In our analysis, we have found larger number of duplications: 5268 duplications were found at the Osteichthyes ceancestor which represents the early vertebrates period, and 1226 duplications were gained at the Teleostei ceancestor which represents the early fish period. In addition, we have also identified lots of fish specific duplications, with duplications in zebrafish the largest. But as our analysis is more fine-grained, we also found large number of duplications in Percomorphaceae ceancestor, a subclass of the ray-finned fish, and losses in clades like Ovalentaria. They found relatively small number of duplicates for chicken. Similarly, we found chicken to have small number of duplicates compared with other vertebraes, and wild turkey (MELGA, Meleagris gallopavo) has an even smaller scale of duplications. In addition, we find the Sauria clade of the amniotes to have more gene losses compared with the sister clade Mammalia.

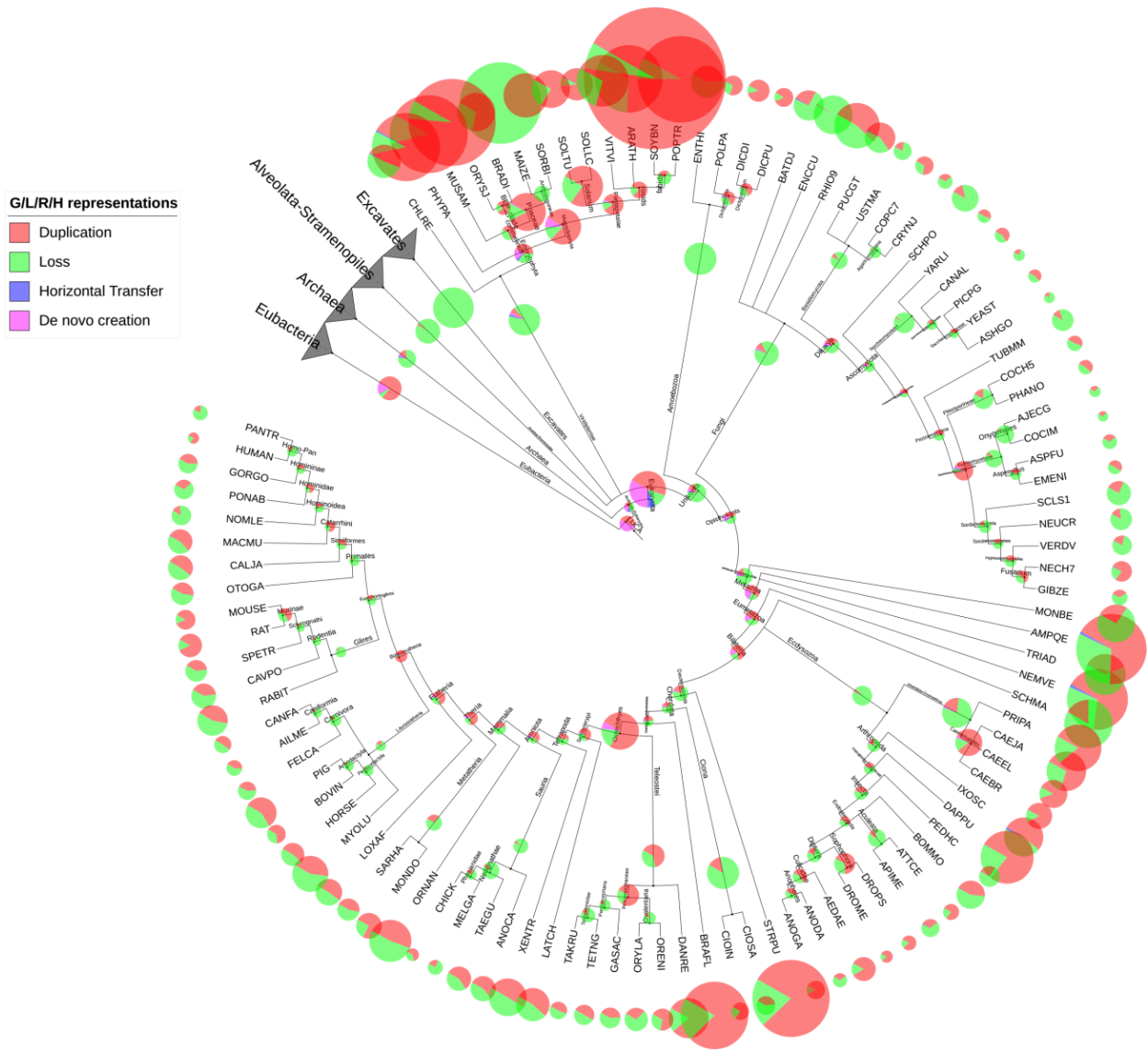

*Figure 2. Evolutionary history of gene gains and losses mapped to reference species tree: SuperPANTHER*

*This figure shows the evolutionary history of gene gains and losses mapped to the SuperPANTHER. For each internal node and each leaf species, there is a pie chart illustrating the relative percentage of the gene copies gained by duplication (red), by de novo creation (magenta) and by horizontal transfer (purple) and the lost gene copies (green). The size of the pie charts is proportional to the total number of gene copies gained and lost. In this figure, four taxa are collapsed as the species in these taxa may not be fully representative.*

## Gene duplications in the human lineage

Gu Xu *et al.* estimated the age of 1739 human gene duplications from 749 vertebrate gene families using an approximate molecular clock method (Gu *et al*. 2002). They find the duplications to be enriched at 3 waves, wave I occurred after mammalian radiation, wave II occurred at an early stage of vertebrate evolution and wave III duplication events took place during metazoan evolution or earlier. Here we plotted the gene evolutionary events in each ceancestor in the evolutionary path leading to human in Figure 3. The figure shows "hot periods" in the evolutionary history leading to human beings, when genome sizes are expanded both greatly and rapidly. The 2 most significant "hot periods" are at the ceancestor of all Eukaryotes and the ceancestor of all vertebrates. We could also find additional 2 periods of rapid gene duplications but with a smaller scale, the first at the emergence of early animals, the second at the emergence of placental mammals and marsupials. Our results are roughly consistent with Gu Xu *et al*'s 3 wave pattern, but provides stronger support for whole genome duplications (WGDs) before divergence of vertebrates and possible WGDs before divergence of Eukaryotes.

## Gene duplications and the extinction events

By relating the great extinction events in geological history and estimated age of ceancestors, we had an interesting founding that large scale gene duplications usually happen after great extinctions in geological time scale.

The great oxygenation event (GOE) was the appearance of dioxygen in Earth's atmosphere produced by photosynthesis of ocenic cyanobacteria (Pinti 2011). It was conferred to have happened around 2.45 billion years ago. Following the great oxygenation event was a glaciation periold named Huronian glaciation that extended from 2.4 billion to 2.1 billion years (Tang and Chen 2013). The hypothesis is that increased atmospheric oxygen combined with the methane to form carbon dioxide and water, which does not retain heat as well as methane does. Molecular phylogenetic reconstruction of the Eukaryotic ceancestoris around 2.1 billion years ago (Hedges, et al. 2004). Fossil recored is consistent with this: *Grypania* is an early, tube-shaped fossil of an eukaryotic alga, it has been dated as far as 2.1 billion year ago (Knoll, et al. 2006). It may not be a conincidence that the Eukaryotic ceancestorevolved after the Huronian gaciation. It is our hypotheis that the drastic change in the envrionment may form a favorable hatch for the emergence of ancient Eukaryote species with abundance of extra genes. The environment change may extinct the vast majory of previously existing species, creating a favorable "vacumm" which is less competitive for the new species to survive with extra amount of genes. Some extant Archae continue to prosper in extreme habitats like hot springs and salt lakes (Chaban, et al. 2006). Besides we think the Eukaryogenesis process may have happened fairly quickly during the Huronian glaciation to adapt to the new environment, in contrast to the comment of Booth *et al*: "take Eukaryotes 2 billion years to evolve" (Booth and Doolittle 2015).

End-Ediacaran extinction suggests a mass extinction of acritarchs and sudden disappearnce of Ediacara biota 542 million years ago (Darroch, et al. 2015). It is followed by the Cambrian explosion which occurs 541 million years ago in the Cambrian period and lasts about 20-25 million years (Conway Morris 2000). The Cambrian explosion has witness the appearnce of most modern metazoan phyla. This may be associated with the large number of duplications and de novo creations in Metazoa, Eumetazoa, Bilateria branches. Noteblely, the earliest chordate fossil found is "Haikouella lanceolata" from about 521 millions years ago in Chengjiang fauna China (Chen, et al. 1999). Interstingly, the most succesful chordates nowadays, the Osteichthyes (bony fish) didn't evolve until another great extinction events 100 million years later. The Ordovician–Silurian extinction events are dated in the interval of 455-430 ma ago (Barash 2014). Almost all major taxonomic groups were affected during this extinction event. Extinction was global during this period, eliminating 49-60% of marine genera and nearly 85% of marine species.Interestingly, the oldest known fossils of Osteichthyes(bony fish) are about 420 million years, right after the ordovician-silurian extinctions. The popular 2R hypothesis states that 2 rounds of whole genome duplications have contributed to the emergence of early vertebrates (Ohno 1970). Our data show extra large amount of ancient duplications at the specific time period of Osteichthyes is consistent with the 2R hypothesis. It is our hypothesis that the great extinction has created favorable environment for the survived chordates to undergo 2 rounds of wholge genome duplications which provide extra amount of gene material to allow further evolution.

## Gene evolutionary events in insects

Stefan Wyder *et al.* compared the genes of five vertebrates (human, mouse, opossum, chicken, and pufferfish) and five insects (fruitfly, malaria mosquito, dengu/yellow fever mosquito, honeybee and red flour beetle) to identify gene losses in vertebrates and insects (Wyder *et al*. 2007). In our analysis, 9 insects were included, but we missed red flour beetle was not included. Detailed results for the gene gains and losses in the insect's phyla are shown in Figure 4.a. For a more direct comparison, we built a sub species tree for APIME (honeybee), DROME (fruit fly), AEDAE (dengu/yellow fever mosquito) and ANOGA (malaria mosquito) and extracted the gene loss information from the SuperPANTHER reference species tree. We compared with Stefan Wyder's results, and find several similar patterns in both researches: (1) the honeybee (APIME) branch has experienced the largest number of gene losses; (2) malaria mosquito (ANOGA) has more gene losses than dengu/yellow fever mosquito (AEDAE); (3) The total number of gene losses in fruit fly is larger than that of ANOGA or AEDAE since divergence their ceancestorDiptera; (4) the Diptera branch has experienced more gene losses than Anopheles branch. The detailed comparison of gene losses in the insect phyla is shown in Figure 4.b.

The inference of gene loss is dependent on the species tree structure and the number of species used for comparative analysis. With more species, we could build a more detailed tree with more internal nodes (common ancestors), and more duplications could be inferred at these internal nodes. Thus, we could infer more losses resulting from these duplications. Stephan Wyder et al. observed significantly more frequent gene losses in insects than in vertebrates. However, our results do not support such a pattern. The total number of losses in the human lineage from Bilateria (the ceancestorof insects and vertebrates) is 8840, many of the losses are accredited to the duplications at the 19 common ancestors in the lineage from Bilateria to Human. In Stephan Wyder et al.'s analysis, only 5 vertebrates were used, forming only 4 common ancestors in the same lineage. Thus, more gene losses in vertebrates were inferred correctly by our analysis. We see there are a large number of gene losses in each of the 3 descendants of

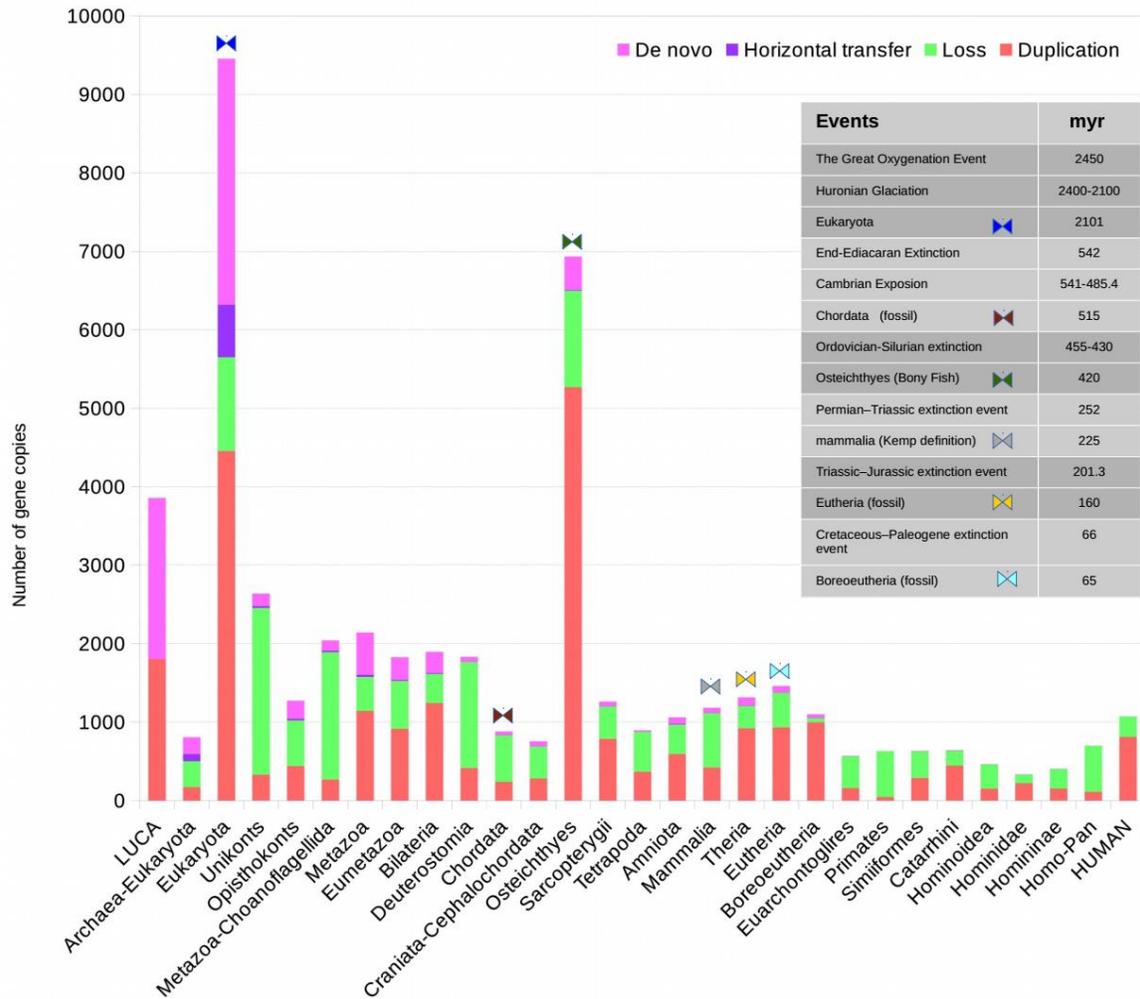

**Figure 3. History of gene gains and losses along the human lineage**

*In this figure, we've plotted the number of gene gains and losses for each ceancestorin the human lineage. The left y axis is the absolute number of gene copies; The x axis is the common ancestors from the least ceancestor(LUCA) on the left to the ceancestorof Human and Chimpanzee (Homo-Pan) on the right. For each common ancestor, a stacked bar depicts the number of gene gains by de novo creation (magenta), horizontal transfer(purple) and Duplication (red), and gene losses(green).     From this picture, we see the rate of gene gains and losses vary greatly at different evolutionary periods. There is a considerably large number of duplications and de novo creations at LUCA and Eukaryota branches which represent the earliest life forms on Earth billions years ago. On the right of the figure is a table which summarizes some great extinction events and fossil records of several common ancestors which have large number of duplications.*

Endopterygota: BOMMO, Aculeata and Dipetera. Diptera has 18 new genes created, and gained 4 copies of genes by horizontal transfer, showing an elevated evolutionary speed in Diptera clade. Leafcutter ant (Atta cephalotes ATTCE) and Honeybee (Apis mellifera, APIME) both belong to th Aculeata clade, but they have distinct evolutionary patterns after their divergence. Leafcutter ant gained more genes than lost, while honeybee lost more genes than gained, but has gained 55 copies of genes via horizontal transfer. Similar pattern was found in descendants of Anopheles: ANOGA (Anopheles gambiae, African malaria mosquito) has gained more gene copies than lost while ANODA (Anopheles darlingi, Mosquito) has lost more than gained.

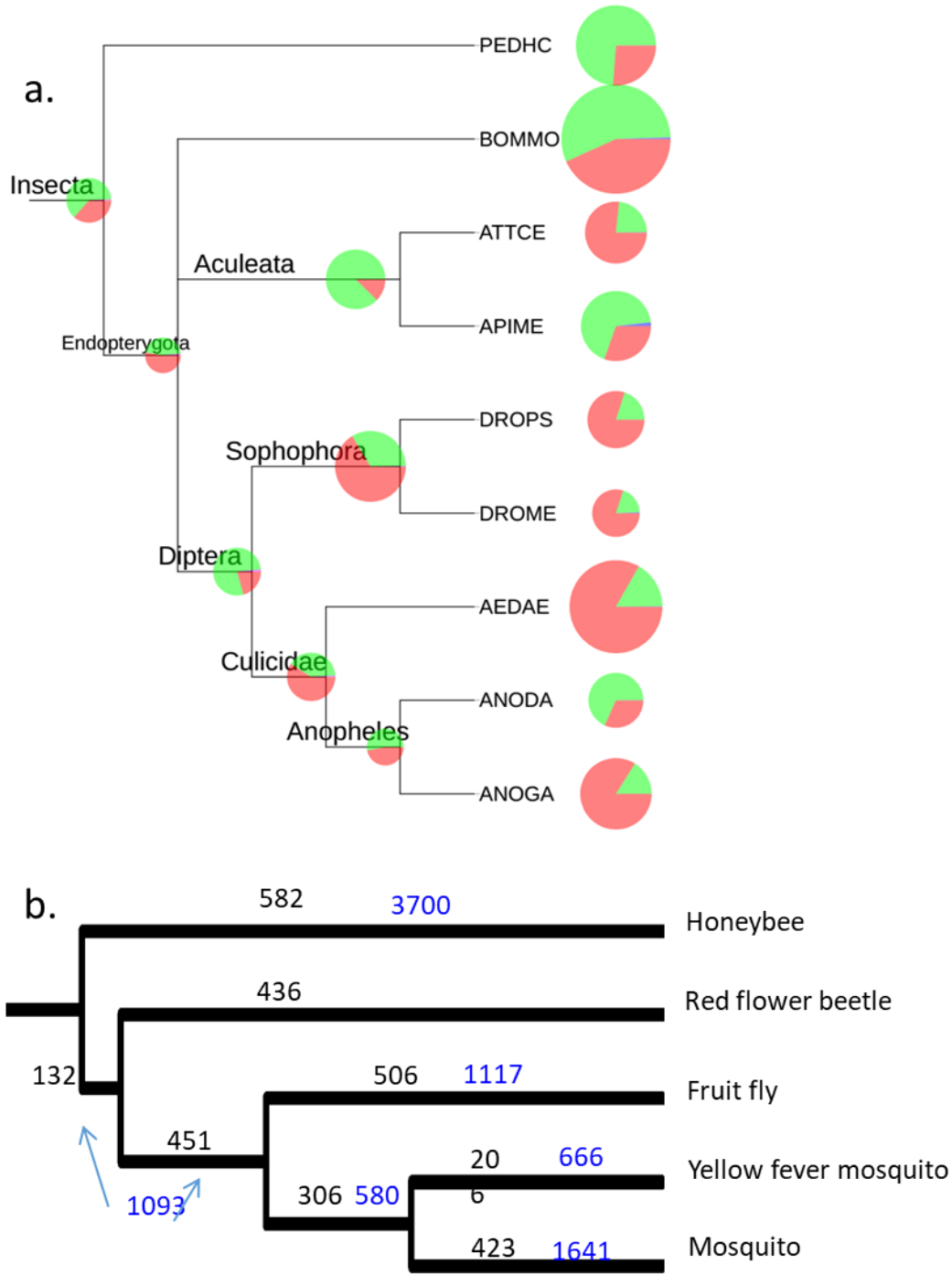

Figure 4. Evolutionary events in insect phyla

*a. The evolutionary history of gene gains and losses for insects phyla in SuperPANTHER 1.0*

*This is our result for insect's phyla. Each branch in this big tree has 4 numbers, duplication, loss, horizontal transfer and root. We see large number of gains and losses in terminal branches and smaller numbers in internal branches. Some horizontal transferes in honeybee, and silkworm. We only see a few creations of new gene families before Diptera and Culicidae.*

*b. Comparisons of gene losses for insect phyla with Wyder el al.'s research*

*Wyder et al. performed genomic reconstructions with honeybee, red flower beetle, fruit fly, yellow fever mosquito and mosquito and have only looked at losses. For comparison, we extracted results from our reconstructions. The black numbers are their estimated losses, and the blue numbers are our estimates. The absolute numbers are quite different. Our numbers are much larger than theirs, but it is expected because we've analyzed more gene families and we've compared with more species. But the ratios are in reasonable range.*

## Gene evolutionary events in Eubacteria

Pere Puigbo *et al.* have performed a comprehensive study of 34 groups of bacteria and one group of archaea (Puigbo *et al.* 2014). Each group is a collection of align-able tight genome clusters (ATGCs). Unlike our analysis, they not only consider gains and losses of genes, but also expansion and reduction of genes among prokaryotes members within a group. As only one representative genome for each species is used in SuperPANTHER, we do not have the expansion and reduction results. But SuperPANTHER 1.0 has 69 different bacteria species, thus our analysis could be more detailed for between genome clusters changes. Detailed results of gene evolutionary events in Eubacteria are listed in Figure 5.

Pere Puigbo *et al.* have found gene loss to be the dominant process in prokaryote genome evolution and is nearly threefold higher than the gain rate. The rates of genome changes are remarkably high for the genome size of bacteria and archaea. Only several groups of bacteria like Enterobacter, Klebsiella, Campylobacter and Listeria have more gains than losses. Legionella and Corynebacterium have extensive gene losses. The Enterobacteriaceae and Campylobacterales clades include both gainers and loser ATGCs. Consistently, we observe extensive gene losses at nearly all bacteria branches, several folds of gene gains. Gain of genes was mainly through gene duplications, and less commonly through horizontal gene transfers. De novo creation of new genes is very rare in nearly all branches, except small numbers in Proteobacteria, Gammaproteobactia, Actinomycetales, Firmicutes and Cyanobacteria branches. Even in these branches, much more genes were gained by duplications or horizontal gene transfers. 10 bacteria species have more gains than losses, making them distinct from the rest 59 species. We also find more gains than losses in Proteobacteria ceancestor. Pere Puigbo *et al.* grouped many member of the Enterobacteriaceae clade like Citrobacter koser, Enterobacter, Escherichia Coli, Salmonella and Shigella together as one ATGC. In our study, descendants of Enterobacteriaceae branch including (*Shigella*), SALTY (*Salmonella*), ECOLI (*Escherichia Coli*) and YERPE (*Yersinia pestis*) were analyzed separately. YERPE (Yersinia pestis) has more duplications than losses, while BUCAI (*Buchnera aphidicola*) has almost no gene duplications.

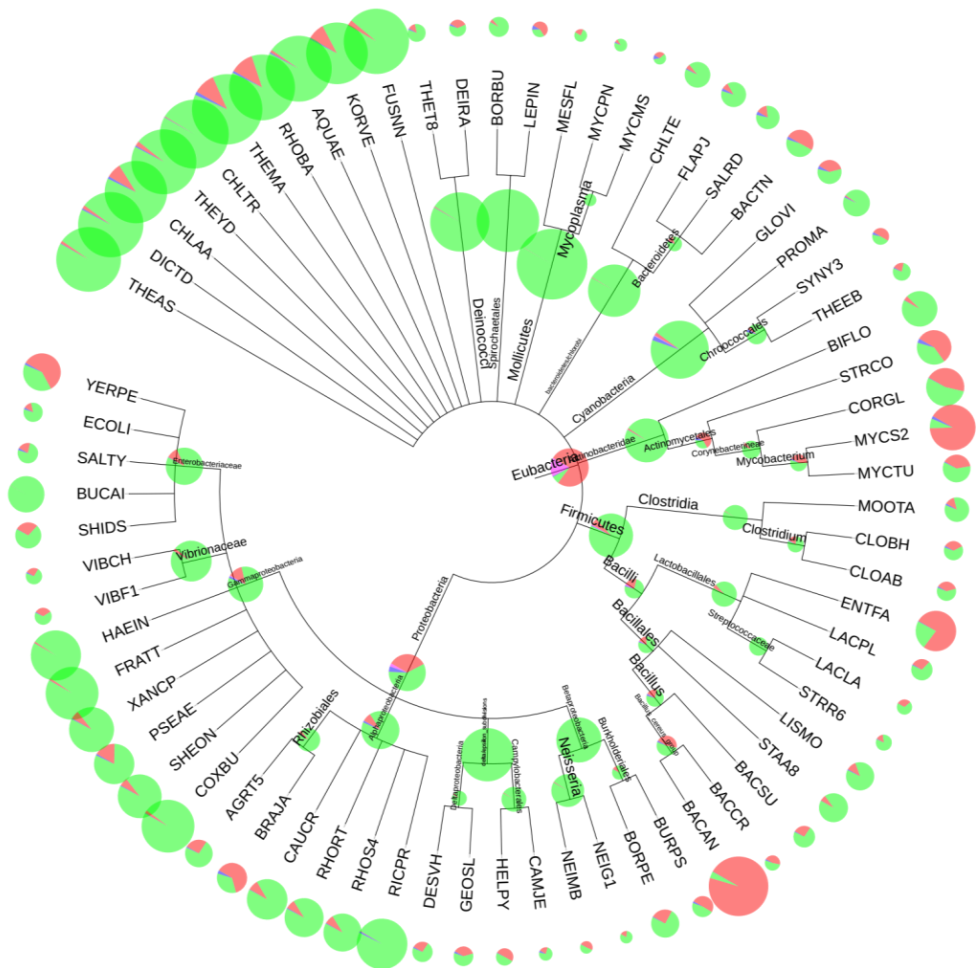

*Figure 5 gene gains and losses in Bacteria phyla*

*Gene losses dominate the ceancestors and extant species with a few exceptions. The Eubacteria ceancestor has many duplications and de novo creation of gene families, the proteobacteria ceancestor also sees duplications and new gene families. Details are described in the results part.*

Both our method and Pere Puigbo *et al.*'s research have poor representation of all bacterial species: 34 and 69 different bacterial species seperately; but there are about 15979 known species of bacteria and archaea (Parte 2014). Besides, the attempts to estimate the true number of bacterial diversity have ranged from $10^7$ to $10^9$ total species (Curtis *et al*. 2002; Schloss and Handelsman 2004). The small sample of bacterial genomes would very likely lead to unbalanced conclusions in evolutionary history of gene gains and losses in the bacterial clade.

Another source of artifact may come from the model we used for reconstruction of gene gain, inheritance and loss. The Eubacteria branch has a total of 3242 duplicates, 682 new gene families, and 21 horizontal transfers. The total number of genes before the bacterial radiation at Eubacteria would be 7688 based on our model. Although the number is several folds larger than the total number of genes in extant bacteria genomes, we may still have underestimated it due to the limited number of bacteria species in our study. The pangenome of *Escherichia coli* contains more than 20,000 genes (Medini *et al*. 2005). According the current model, the enormous number of total genes in the Eubacteria ceancestor could lead to greatly inflated amount of losses in descendants of Eubacteria. In our analysis, large number of genes are inferred to be lost in descendant clades of Eubacteria: more than 2600 gene copies are lost at CHLAA (*Chloroflexus aurantiacus*), DICTD (*Dictyoglomus turgidum*)and Cyanobacteria branches. The large number of estimated gene losses in these branches, together with large number of gains at the Eubacteria ceancestor are probably artifacts. In the species tree, Eubacteria has 18 direct descendants. If one or more duplicate gene exists in 2 or more of these descendants (let's call them group A descendants for simplicity of explanation), we estimate the duplication event happens before the divergence of Eubacteria which would imply a duplicated copy of gene in all the 18 direct descendants. Thus the descendants that do not have the duplicated copy would be inferred to have lost it. Thus, we may observe an artifact in inflation of the duplications at Eubacteria and losses in its descendants.

## Horizontal transfers among major phyla

Horizontal gene transfer (HGT) is the transfer of DNA between diverge organism. Horizontal transfers have found among major phyla: within bacteria (Knoppel *et al*. 2014; Bonham *et al*. 2017), within fungi (Fitzpatrick 2012), within plants (Gao *et al*. 2014), between bacteria and animals (Dunning Hotopp 2011; Husnik *et al*. 2013; Sun *et al*. 2015), from a Mosquito Host to Gut Fungi (Wang *et al*. 2016).

To our knowledge, this analysis is a first attempt to identify horizontal gene transfers across the entire tree of life. We have found horizontal transfers mostly within Eubacteria, which consist about 1/3 of all gained genes and are significantly abundant than de novo creation of genes. Many horizontal transfers are also found in Fungi. Details are shown in Figure 6. More horizontal transfers were found in the lineage of multicellular fungus (like ASPFU, *Aspergillus fumigatus*) than the lineage of unicellular fungus (like YEAST, *Saccharomyces cerevisiae*). The results are consistent with the recent research of Gergely J szollosi *et al*'s comparative analysis of 32 fungus species (Szollosi *et al*. 2015).

Lots of horizontal transfers are also found within Archaea, Alveolata-Stramenopiles, Excavates and Amoebozoa. For example, METAC (*Methanosarcina acetivorans,* an archaea) has gained 231 copies of genes after its divergence from Methanomicrobia, APIME (*Apis mellifera*) has gained 55 copies of genes after its divergence from Aculeata.

Horizontal transfers may have played a crucial role in the evolution of early living forms. We estimate 88 horizontal transfers to Archaea-Eukaryota, 654 horizontal transfers to Eukaryota, and 114 horizontal transfers to Viridiplantae (green plants) after divergence of Eukaryota. Most of the transfers to the Eukaryota branch are from proteobacterial and Cyanobacteria lineages. This finding may provide evidence for the hypothesis that most of these genes originated from the endosymbiosis event from bacteria leading to the mitochondrion. In addition, most of the transfers to the Viridiplantae branch are from cyanobacterial lineages, which provides support for chloroplast endosymbiosis (McFadden and van Dooren). Horizontal transfers are rarely observed after Deuterostomia ceancestor. As exception, we observe 15 horizontal transfers in ORNAN (*Ornithorhynchus anatinus*), 19 in ANOCA (*Anolis carolinensis*), 27 in MACMU (*Anolis carolinensis*). The detailed results of all horizontal transfer events are listed in Supplemental file 6.

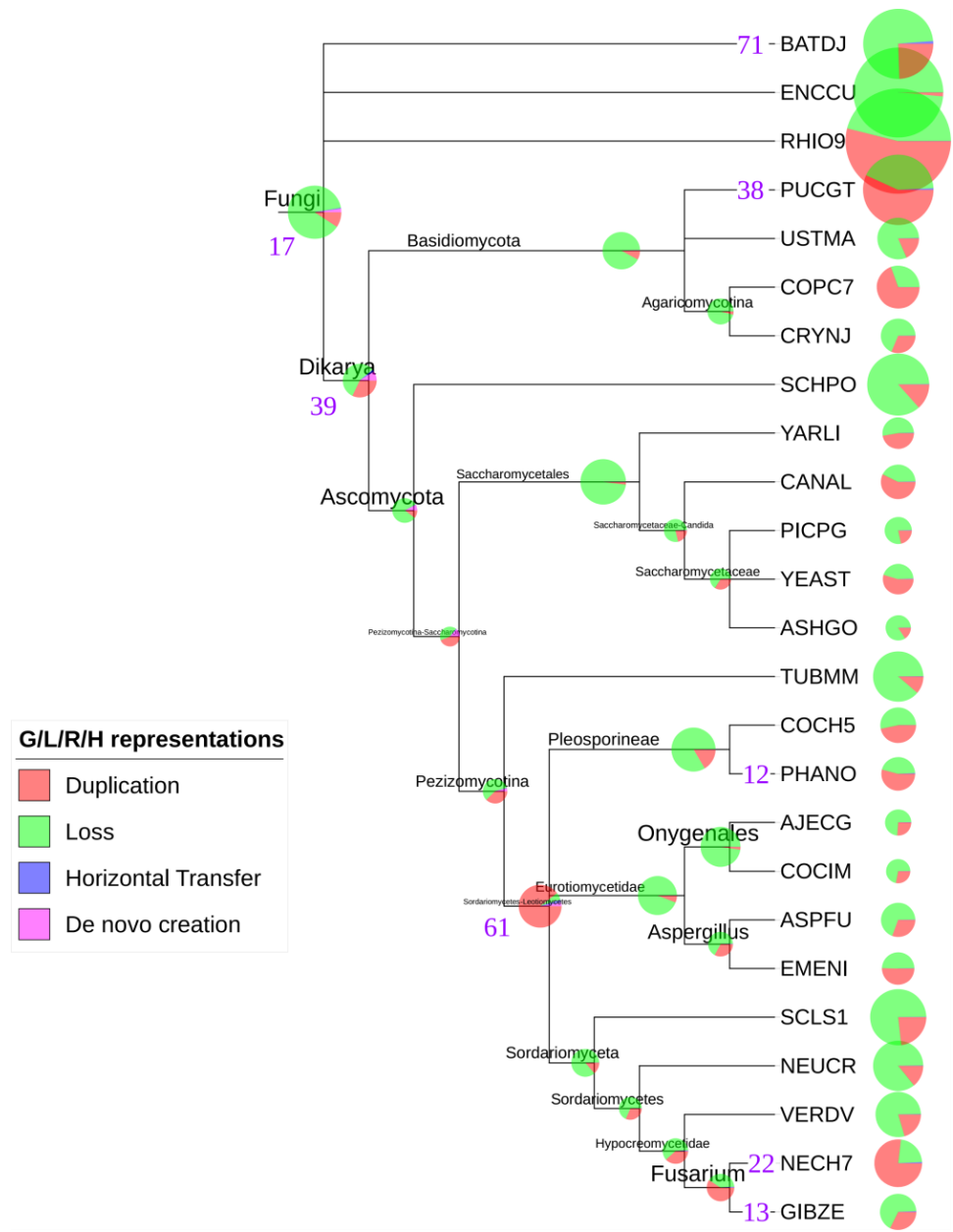

**Figure 6 Horizontal gene transfers in the Fungi phyla**

*To better illustrate the number of horizontal transfers in Fungi, we add the absolute number in purple below the branch name (like 39 transfers to Dikarya) or before the species name (like 71 transfers to BATDJ). All the internal nodes and leaf species without the number label have fewer than 10 horiziontal transfers.*

## Novel patterns, hypothesis and artifacts

As this study is the first attempt to reconstruct the gene evolution events among all major phyla of life, we have found many novel patterns that had not been reported in previous studies. The accurate construction of phylogenetic gene tree has been notoriously difficult, thus imperfections in

SuperPANTHER1.0 would inevitably lead to problems of the gene evolution reconstructions. Besides, the phylogeny of the Eukaryotic least common ancestor's (LECA) is not fully resolved, thus we have to use non-bifurcating tree structure for LECA's descendant clades. This would lead to inflations of the gene gains in LECA, and subsequent inflations of gene losses in descendant clades. It is worth deeper discussion as to whether the evolutionary reconstructions reflect the accurate evolutionary history. As the SuperPANTHER is balanced towards human and vertebrate genomes, the other clades may not be fully represented. However, it is evident from comparisons with previous studies with fewer representative species that our analysis depicts a much more fined grained evolutionary history while the major gains and losses remain largely consistent.

The green plants have large number of gene gains ever since the viridiplantae ceancestor, represented in duplications in major ceancestors as well the species specific duplications. The ceancestor of land plants (Embryophyta) and the ceancestor of flowering plants (Magnoliophyta) have a large number of de novo creations. In contrast, other plant ceancestors have few de novo creations, but instead much larger number of duplications. This pattern may support the hypothesis that novo creation of genes is expensive in evolution that only happen in circumstances when essential new functions cannot be achieved by duplicating or re-organizing the existing genetic material.

Amoebozoa is a monophyletic clade of amoeboid protist, a sister group to Opisthokonda which contains both animals and fungi. In our analysis, Amoebozoa is a distinct clade of protists whose genomes have greatly reduced in size after its divergence from the Eukaryotic ceancestor. In addition, the protists in our analysis distinct with each other by species specific duplications. Besides, we have found considerably large number of duplications in "outgroup" species, like Choanoflagellate (MONBE, Monosiga brevicollis) which is a group of free-living unicellular and colonial flagellate eukaryotes considered to be the closest living relatives of the animals, sea urchin (STRUPU, Strongylocentrotus purpuratus) and florida lancelet (BRAFL, Branchiostoma floridae), a fish-like marine chordate which is relative of the vertebrates. The large number of gene event changes contrasts with the fossil findings that the appearance of these species remained consistent for millions of years and the common belief that their genomes have also remained constant. However, the imbalance of the phylogenetic tree towards human and vertebrate genes may have inflated the genomic changes in these species.

There are also patterns that we do not how to properly explain. For example, the ratio of gene gains vs gene losses is consistent in several nearby ceancestors in the human lineage. The pairs of Eumetazoa and Bilateria ceancestors, the Deuterostomia and chordata ceancestors and the Amniota, Sarcopterygii, Theria and Eutheria have similar pattern of gene gains and losses. We also noticed large number of gene losses in the Sauria (lizards) clade as well as the Laurasiatheria clade. The classification of the later is based on DNA sequence analyses and retrotransposon presence/absence data. The name comes from the theory that these mammals evolved on the supercontinent of Laurasia, after it split from Gondwana when Pangaea broke up. It would be interesting to further explore why Laurasiatheria clade has larger number of gene losses compared with the sister group Euarchontoglires (Supraprimates) of which human beings belong to.

## Conclusion

Here, we have constructed and analyzed over 10,000 gene family trees to reconstruct the gene content of ancestral genomes at an unprecedented scale, covering hundreds of genomes across all domains of life. We find that our results are largely consistent with earlier, less complete comparative studies on specific lineages such as the vertebrates. We also find that the rate of gene gain varies widely along the evolutionary path leading to human beings. There are roughly 3 periods of rapid gene duplications in the human lineage, the first at the emergence of early animals, the second at the emergence of early vertebrates and the third at the emergence of placental mammals and marsupials. By relating the great extinction events in geological history and estimated age of ceancestors, we had an interesting founding that large scale gene duplications usually happen after great extinctions. The Osteichthyes and Teleostei ceancestors have very large numbers of duplications which provide evidence for the 2-R whole genome duplications and the fish specific genome duplications. We found horizontal transfers at nearly all major domains of life, including bacteria, fungi, archaea, vertebrates and others, many of which haven't been described before in previous literatures. In summary, our study has yielded a complex and detailed evolutionary history of gene gains and losses since the universal least common ancestor to extant species. Our results may reveal insights for researchers who are interested in gene evolutionary histories in other lineages as well. Besides, the algorithm of gene evolutionary history reconstruction and data visualization method are very general, they may be utilized for analysis of protein classification databases.

## Data and code availability

All software for inferring ancestral gene content from gene trees is available at https://github.com/haimingt/genome_history_inference.

## Acknowledgements

This work was supported by the National Science Foundation [grant number 1458808] and the National Human Genome Research Institute of the National Institutes of Health [award number U41HG002273].